\documentclass[aps,pra,twocolumn,floatfix,superscriptaddress,10pt]{revtex4-2}
\usepackage{mathbbol}
\usepackage{amsmath,amssymb}
\usepackage{tikz}
\usepackage{centernot}
\usepackage[utf8]{inputenc}
\usepackage[english]{babel}
\usepackage{amsfonts}
\usepackage{amsthm}
\usepackage{mathrsfs}
\usepackage{mathtools}
\usepackage[nice]{nicefrac}
\usepackage{amsmath}
\usepackage{xfrac}
\usepackage{tabularx}
\usepackage{physics,subfigure}
\usepackage{tikz}
\usetikzlibrary{positioning}
\usepackage{tkz-berge}
\usetikzlibrary{arrows}
\usetikzlibrary{graphs,graphs.standard,fit,shapes.geometric,calc,hobby}
\begin{document}
\title{Perfect Chiral Quantum Routing}
\author{Simone Cavazzoni}
\email{simone.cavazzoni@unimore.it}
\affiliation{Dipartimento di Scienze Fisiche, Informatiche e Matematiche,  Universit\`{a} di Modena e Reggio Emilia, I-41125 Modena, Italy}
\author{Giovanni Ragazzi}
\email{giovanni.ragazzi@unimore.it}
\affiliation{Dipartimento di Scienze Fisiche, Informatiche e Matematiche,  Universit\`{a} di Modena e Reggio Emilia, I-41125 Modena, Italy}
\author{Paolo Bordone}
\email{paolo.bordone@unimore.it}
\affiliation{Dipartimento di Scienze Fisiche, Informatiche e Matematiche, Universit\`{a} di Modena e Reggio Emilia, I-41125 Modena, Italy}
\affiliation{Centro S3, CNR-Istituto di Nanoscienze, I-41125 Modena, Italy}
\author{Matteo G. A. Paris}
\email{matteo.paris@fisica.unimi.it}
\affiliation{Quantum Technology Lab, Dipartimento di Fisica {\em Aldo Pontremoli}, Universit\`{a} degli Studi di Milano, I-20133 Milano, Italy}
\affiliation{INFN, Sezione di Milano, I-20133 Milano, Italy}
\date{\today}
\begin{abstract}
Routing classical and quantum information is a fundamental task for quantum information technologies and processes. Here, we  consider information 
encoded in the position of a quantum walker on a graph, and design an 
optimal structure to achieve perfect quantum routing exploiting chirality and 
weighting of the edges. The topology, termed the {\em Lily Graph}, enables perfect (i.e., with fidelity one) and robust routing of classical 
(localized) or quantum (superposition) states of the walker to $n$ different, orthogonal, spatial regions of the graph, corresponding to the $n$ possible outputs of the device. The routing time is independent of the input signal and the number of outputs, making our scheme a robust and scalable solution for all quantum networks.
\end{abstract}
\maketitle
\section{Introduction} 
Routing of information is a fundamental procedure, 
crucial for any classical or quantum information protocol \cite{yung2005perfect,zwick2011robustness,kendon2011perfect,Shomroni2014903,Lu2014,nikolopoulos2014quantum,Caleffi201722299}. Networks that exchange information across different devices have to be equipped with a mechanism for selecting paths through the network itself \cite{sazim2015retrieving,pant2019routing,bottarelli2023quantum,ren2022nonreciprocal}. Since quantum mechanics relies on probability amplitudes and interference patterns, the main challenge of this approach is to create a universal scheme that takes an initial state $\ket{\psi_{0}}$ and evolves it in different possible orthogonal states $\ket{\psi_{f}}$, according to some tunable parameters of the system, with unit probability at a time $t^{*}$. The orthogonality condition guarantees the perfect discriminability of the different $\ket{\psi_{f}}$. This condition can be implemented associating to $\ket{\psi_{0}}$ different states $\ket{\psi_{f}}$ belonging to separated spatial regions. In the following, we will refer to $\ket{\psi_{0}}$ as the \textit{input} state and to $\ket{\psi_{f}}$ as an \textit{output} state. 
\par
In this work, we consider information encoded in the position of a 
quantum walker on a graph \cite{underwood2012bose,PhysRevLett.129.160502}. Routing this information means to drive the walker from a given input state to $n$ possible output states by tuning a system parameter. Classical information corresponds to preparing the walker in a localized state, while quantum information may be encoded using superpositions of at least two localized states. We look for an optimal graph topology that ensures the routing of both classical and quantum information with unit fidelity in the shortest possible time. Indeed, by exploiting edge weighting and chirality, we have designed an optimal structure ensuring perfect routing. More specifically, using chirality alone, we achieve nearly optimal routing, while the combined use of chirality and edge weighting enables us to create a perfect router. This router achieves perfect (fidelity one) and robust routing in a time independent of the input signal and the number of outputs.
\par
The article is structured as follows. In Sec.\ref{sec:IQR}-\ref{sec:DR} we introduce the theoretical background, from the router mathematical description and requirements to the dimensionality reduction method adopted for the analysis. In Sec.\ref{sec:TLG}-\ref{sec:PRUCAEW} we depict the topology that supports the perfect chiral quantum routing, presenting the analytical results obtained using phases and edge weighting. In Sec.\ref{sec:QR} we generalize our result to the qudit routing, and then to a universal and scalable routing of quantum information procedure, proposing in Sec.\ref{sec:PI} also possible physical implementations of the protocol. Finally in Sec.\ref{sec:C} we draw our conclusions.
\section{Ideal Quantum Router} 
\label{sec:IQR}
An \textit{ideal quantum router} is a system in which the time evolution, setting $\hbar=1$, is governed by
\begin{equation}
    \label{eq:ideal quantum routing}
    e^{-i H\,t^{*}} \ket{\psi_{0}} = \ket{\psi_{f}},
\end{equation}
i.e. the time evolution operator acts as a projector $\ket{\psi_{f}} \bra{\psi_{0}}$. An ideal quantum router should be able to carry not only classical information (for $\ket{\psi_0}$ and $\ket{\psi_f}$ being localized states) but quantum information as well (requiring $\ket{\psi_0}$ and $\ket{\psi_f}$ to be superpositions of localized states). Assuming that the system consists of a quantum walker on a graph, the time evolution of the system takes place in a position Hilbert space $\mathscr{H} = \operatorname{span}\{ \vert x \rangle \}$. The states $\vert x \rangle$ are the sites of a $\mathcal{N}$ dimensional discrete space (i.e. a graph $\mathcal{G} \left( \mathcal{V},\mathcal{E} \right)$ of vertices $\mathcal{V}$ and edges $\mathcal{E}$). The discrete topology itself defines the adjacency matrix of the graph $A$, whose element $A_{jk}$ are defined by the edges as
\begin{equation}
\label{eq:adjacency_matrix}
A_{jk} = \begin{cases}
-1 & \text{if  $j \neq k$ and $(j, k) \in \mathcal{E}$,}\\
0 & \text{otherwise.} 
\end{cases}
\end{equation}
which is a valid generator for the dynamics of a Continuous Time Quantum Walk (CTQW). Nonetheless, 
the quantumness of the system allows us to consider also chiral 
adjacency matrices \cite{Lu2016,frigerio2021generalized,Khalique2021,cavazzoni2022perturbed}
\begin{equation}
\label{eq:adjacency_matrix_chiral}
C_{jk} = \begin{cases}
e^{-i\phi_{jk}} & \text{if  $j \neq k$ and $(j, k) \in \mathcal{E}$,}\\
0 & \text{otherwise,} 
\end{cases}
\end{equation}
defining a set of Hamiltonians generators $ H  = C=C^\dag$ for 
quantum dynamics. Weighting the edges of the topology is also 
possible and corresponds to tuning the moduli of the elements 
of the Hamiltonian, e.g. setting $|H_{jk}|=\beta \in \mathbb{R}_+$.
\section{Dimensionality Reduction} 
\label{sec:DR}
In a quantum router, the quantities of interest are the probability amplitudes at  particular vertices of the graph. Exploiting symmetries of the discrete structure, a dimensionality reduction 
method provides a tool to define an effective topology for the time evolution of the  system \cite{caruso2009highly}. Considering a Taylor expansion of the quantum time evolution operator \cite{novo2015systematic}, the probability amplitude at a vertex $\ket{\zeta}$ can be expressed as
\begin{align}
\label{eq:dim_red}
\bra{\zeta}e^{-i H t}\ket{\psi_{0}}&=\sum_{k=0}^\infty \frac{(-it)^k}{k!}\bra{\zeta} H ^k\ket{\psi_0}\nonumber\\
&=\bra{\zeta}e^{-i \widetilde{H} t}\ket{\widetilde{\psi}_0}\,,
\end{align}
where  $ {\cal P}H{\cal P} =\widetilde{H}$ is a reduced Hamiltonian, $\ket{\widetilde{\psi}_0}= {\cal P} \ket{\psi_0}$ is a reduced state, and 
 ${\cal P} $  is the projector onto the Krylov subspace, 
 which itself is defined as
\begin{equation}
\mathcal{K}( H ,\ket{\zeta}) = \operatorname{span}(\lbrace  H ^k \ket{\zeta} \mid k \in \mathbb{N}_0\rbrace)\,,
\label{eq:Krylov_subspace}
\end{equation}
ensuring that $\ket{\widetilde{\zeta}}= {\cal P} \ket{\zeta}=\ket{\zeta}$. Clearly, $\dim \mathcal{K}(  H ,\ket{\zeta}) \leq \dim \mathscr{H}=\mathcal{N}$, as $\mathcal{K}(  H ,\ket{\zeta}) \subseteq \mathscr{H}$. The orthonormal basis for the subspace $\mathcal{K}( H ,\ket{\zeta})$ (i.e. $\{\ket{e_1},\ldots,\ket{e_m}\}$) is built iteratively according to a Gram-Schmidt like orthonormalization procedure.

\begin{align}
\label{eq:Gram_Schmidt}
\ket{u_{k}} = \ket{w_{k}} - \sum_{j=1}^{k-1} \bra{e_{j}}\ket{w_{k}}\ket{e_{j}}  \; \xrightarrow{} \; \ket{e_{k}} = \frac{\ket{u_{k}}}{\sqrt{\abs{u_{k}}^{2}}},
\end{align}
with $\ket{e_{1}}=\ket{\zeta}$ and $ H \ket{e_{k-1}}=\ket{w_{k}}$. The original problem is then mapped onto an equivalent one governed by a tight-binding Hamiltonian with $m$ sites. The subspace obtained by means of the dimensionality reduction method provides a subspace which is relevant for the chiral quantum routing.


\section{The Lily graph} 
\label{sec:TLG}
By engineering the topology and chirality of 
the graph it is possible to achieve ideal quantum routing. To this aim
we put forward the structure, termed the \textit{Lily graph}, depicted in 
Fig.\ref{fig:Lily Graph}. 

\begin{figure}[!ht] 
    \begin{tikzpicture}[thick,scale=0.7, every node/.style={scale=0.7}]
    \tikzset{root node/.style={circle,fill=green,minimum size=0.75cm,inner sep=0pt}}
    \tikzset{two node/.style={circle,fill=magenta,minimum size=0.75cm,inner sep=0pt}}
    \tikzset{chiral node/.style={circle,fill=cyan,minimum size=0.75cm,inner sep=0pt}}
    \tikzset{constructive node/.style={circle,fill=yellow,minimum size=0.75cm,inner sep=0pt}}
    \tikzset{f node/.style={circle,fill=orange,minimum size=0.75cm,inner sep=0pt}}
    
    \tikzset{dots node/.style={circle,fill=white!40,minimum size=0.75cm,inner sep=0pt}}
      \node[f node] (5) {$f$};
      \node[constructive node] (4) [below = 0.75cm of 5]  {$r$};
      \node[chiral node] (3) [below = 2.125cm of 5]  {$k$};
      \node[two node] (2) [below = 3.25cm of 5]  {$2$};
      \node[root node] (1) [below = 4.625cm of 5]  {$1$};
      \node[chiral node] (6) [left = 0.75cm of 3]  {$k$};
      \node[chiral node] (8) [right = 0.75cm of 3]  {$k$};
      \node[constructive node] (10) [left = 2.0cm of 4]  {$l$};     
      \node[constructive node] (11) [right = 2.0cm of 4]  {$l$}; 
      \node[f node] (12) [left = 3.0cm of 5]  {$o$};     
      \node[f node] (13) [right = 3.0cm of 5]  {$o$}; 
      \path[draw=green,thick]
      (1) edge node {} (2);
      \path[draw=magenta,thick]
      (2) edge node {} (3)
      (2) edge node {} (6)
      (2) edge node {} (8);
      \path[draw=cyan,thick]
      (3) edge node {} (4)
      (4) edge node {} (6)
      (4) edge node {} (8);
      \path[draw=yellow,thick]
      (4) edge node {} (5); 
      \path[draw,dotted]
      (10) edge node {} (6)
      (10) edge node {} (3)
      (10) edge node {} (8);
      \path[draw,dotted]
      (11) edge node {} (6)
      (11) edge node {} (3)
      (11) edge node {} (8);
      \path[draw,dotted]
      (10) edge node {} (12)
      (11) edge node {} (13);
    \end{tikzpicture}
    \caption{\textit{Lily Graph} topology -- The \textit{input nodes} ($\ket{1}$ and $\ket{2}$) are green and magenta labeled, the \textit{chiral layer} ($\ket{k}$), composed of $d$ nodes, is blue labeled, the \textit{routing layer} ($\ket{l}$ and $\ket{r}$) composed of $n$ vertices is yellow labeled and the $n$ outputs ($\ket{f}$ and $\ket{o}$) are orange labeled.}
    \label{fig:Lily Graph}
\end{figure}

The network is defined by an input layer 
$\mathcal{I}$ of two nodes $\ket{1}$ and $\ket{2}$ (labeled in green and magenta, respectively), a chiral layer $\mathcal{C}$ of $d$ vertices $\ket{k}$ (blue labeled), a routing layer $\mathcal{R}$, composed of $n$ vertices 
$\ket{l}$ and $\ket{r}$ (yellow labeled) and, finally, and output layer 
${\mathcal O}$ of $n$ outputs $\ket{o}$ and $\ket{f}$ (orange labeled). 
The two integers $d$ and $n$ denote independent degrees of freedom of the topology, and may  assume any value \footnote{strictly speaking the chiral parameter 
must be greater than one ($d>1$) since to create interference the input 
wave-function must be split in at least 2 separated spatial regions.}.
The associated Hamiltonian depends on both $d$ and $n$ and also on the vector of the different phases $\vec{\phi}=\{\phi_1,\phi_2,\dots\}$, i.e. $ H \left( n,d,\vec{\phi} \right)$, and is the sum of the adjacency matrices of the different layers and nodes of the discrete structure, i.e., 

\begin{align}
    \label{eq:Lily_Hamiltonian}
     H  \left( n,d,\vec{\phi} \right) = A_\mathcal{I} + C_{\mathcal{C}} \left( d,\vec{\phi} \right) +C_{\mathcal{R}} \left( d,\vec{\phi},n \right) +A_\mathcal{O} ( n ),
\end{align}
where $A_\mathcal{I}$ refers to the input adjacency matrix, i.e.
\begin{equation}
    \label{eq:input_adjacency}
    A_\mathcal{I} = \ket{1}\bra{2} + \ket{2}\bra{1},
\end{equation}
and is associated to the \textit{input} state $\ket{\psi_{0}}$. The chiral adjacency  reads
\begin{equation}
    \label{eq:deph_adjacency}
    C_{\mathcal{C}}\left( d,\vec{\phi} \right) = \sum_{k \in \mathcal{C}} \left( e^{-i\phi_{k}} \ket{2} \bra{k} + e^{i\phi_{k}} \ket{k}\bra{2} \right),
\end{equation}
where the phases $\phi_{k}$ are the $d$ roots of the unity $\sqrt[\leftroot{-2}\uproot{2}d]{1}$, i.e. $\phi_{k} = 2 k \pi/d$. The routing adjacency $C_{\mathcal{R}}(n,d,\vec{\phi})$ is defined as
\begin{align}
    \label{eq:int_adjacency}
    C_{\mathcal{R}} \left( n,d,\vec{\phi} \right) =& \sum_{k \in \mathcal{C}} \sum_{\substack{l \in \mathcal{R} \\ l\neq r}} \left( \ket{k} \bra{l} + \ket{l}\bra{k} \right) \nonumber \\
    +& \sum_{k \in \mathcal{C}} \left( e^{-i\phi_{k}} \ket{r}\bra{k} + e^{i\phi_{k}} \ket{k}\bra{r} \right).
\end{align}
Finally, the output adjacency $A_{out}(n)$ is given by
\begin{align}
    \label{eq:out_adjacency}
    A_\mathcal{O} \left( n \right) = \sum_{a \in \mathcal{R}} \sum_{b \in \mathcal{O}} \left( \ket{a}\bra{b} + \ket{a} \bra{b} \right).
\end{align}

As input and output state we consider respectively any state in the form

\begin{equation}
    \label{eq:input_state}
    \ket{\psi_{0}} = \alpha\ket{1} + e^{i\gamma}\sqrt{1-\alpha^{2}}\ket{2}
\end{equation}
for the input, and, accordingly,

\begin{equation}
    \label{eq:output_state}
    \ket{\psi_{f}} = \alpha\ket{f} + e^{i\gamma}\sqrt{1-\alpha^{2}}\ket{r}
\end{equation}
for the output, with $\alpha \in [0,1]$ and $\gamma\in\left[0,2\pi\right)$.

\section{Nearly perfect routing using only phases}
The routing phenomena we want to implement are those involving an \textit{input} state (sites $\ket{1}$, $\ket{2}$, or one of their superpositions) driven to separated spatial regions (sites $\ket{r}$, $\ket{f}$ or their superpositions, respectively), where $\ket{r}$ can be any site of the routing layer $\mathcal{R}$ and $\ket{f}$ is its associated output node. 
Choosing the phases of the chiral layer as the roots of the unity 
$\sqrt[\leftroot{-2}\uproot{2}d]{1}$, the property
\begin{equation}
    \label{eq:property_roots_of_unity}
    \sum_{k=1}^{d} e^{i\phi_{k}} = \sum_{k=1}^{d} e^{\frac{i 2 \pi k}{d}} 
    = 0 \;\; \forall d>1
\end{equation}
leads by its own to destructive interference in all the vertices $\ket{l}$ of the routing layer and to a constructive interference only in $\ket{r}$. We can then decide where to send the signal, e.g. to the vertex $\ket{r}$, by removing the phases through its links to the chiral layer. Looking at Fig. 1, this means that the $2 \mapsto k$ links have opposite phases compared to the $k \mapsto r$ ones, whereas all the other links $k \mapsto l$ does not have any associated phase, maintaining the destructive effect of the chiral layer. The reduced Krylov basis guarantees that the time evolution of the system involves only the desired output state(s), since it depends only on a single site of the output (chiral) layer, and not on the others. This means that for every time $t$ the dynamics of the system does not involve the unwanted spatial regions, preventing the system from an undesired routing (see the effective graph in the upper panel of 
Fig. \ref{fig:Effective Lily}). Notice that in order to route the state to a different output $\ket{r^\prime}$, it is sufficient to change the $d$ phases of the $k \mapsto r^\prime$ links.

Applying the Krylov reduction method to the Hamiltonian 
$ H \left( n,d,\vec{\phi} \right)$, starting from the vertex 
$\ket{f}$, as described in Eqs.\eqref{eq:dim_red}-\eqref{eq:Gram_Schmidt}, we obtain the following orthonormal basis $\{\ket{e_{k}}\}$
\begin{align}
    \label{eq:Krylov Basis}
    \ket{e_{1}}=& \ket{f}\,, \quad
    \ket{e_{2}}= \ket{r} \nonumber \\
    \ket{e_{3}}=& \frac{1}{\sqrt{d}}\sum_{k \in \mathcal{C}} e^{i\phi_{k}} \ket{k} \nonumber \\
    \ket{e_{4}}=& \ket{2} \,, \quad
    \ket{e_{5}}= \ket{1}
\end{align}
which is analogous to a basis that can be obtained grouping together 
identically evolving vertices \cite{meyer2015connectivity}, and thus 
providing a valid basis for the dynamics of the system. Moreover, 
thanks to the design of the chiral and routing layers, the effective 
topology, and the dimension of its Krylov representation, is 
independent of the number of outputs $n$. 
\begin{figure}[!ht] 
    \begin{tikzpicture}[thick,scale=0.9, every node/.style={scale=0.9}]
    \tikzset{root node/.style={circle,fill=green,minimum size=0.75cm,inner sep=0pt}}
    \tikzset{two node/.style={circle,fill=magenta,minimum size=0.75cm,inner sep=0pt}}
    \tikzset{chiral node/.style={circle,fill=cyan,minimum size=0.75cm,inner sep=0pt}}
    \tikzset{constructive node/.style={circle,fill=yellow,minimum size=0.75cm,inner sep=0pt}}
    \tikzset{f node/.style={circle,fill=orange,minimum size=0.75cm,inner sep=0pt}}
    
    \tikzset{dots node/.style={circle,fill=white!40,minimum size=0.75cm,inner sep=0pt}}

    \tikzset{name node/.style={circle,fill=white,minimum size=0.75cm,inner sep=0pt}}
    
      \node[f node] (5) {$f$};
      
      \node[constructive node] (4) [below = 0.75cm of 5]  {$r$};
      \node[chiral node] (3) [below left = 2.125cm and 0.5cm of 5]  {$k$};
      \node[two node] (2) [below = 3.25cm of 5]  {$2$};
      \node[root node] (1) [below = 4.625cm of 5]  {$1$};

      \node[chiral node] (6) [left = 0.75cm of 3]  {$k$};
      \node[chiral node] (7) [right = 1.25cm of 3]  {$k$};
      \node[chiral node] (8) [right = 0.75cm of 7]  {$k$};

      \node[dots node] (9) [right = 0.25cm of 3] {$...$};


      \path[draw=green,thick]
      (1) edge node {} (2);
      \path[draw=magenta,thick]
      (2) edge node {} (3)
      (2) edge node {} (6)
      (2) edge node {} (7)
      (2) edge node {} (8);
      \path[draw=cyan,thick]
      (3) edge node {} (4)
      (4) edge node {} (6)
      (4) edge node {} (7)
      (4) edge node {} (8);
      \path[draw=yellow,thick]
      (4) edge node {} (5);

    \end{tikzpicture}
    \bigskip

    \begin{tikzpicture}[thick,scale=0.9, every node/.style={scale=0.9}]
    \tikzset{root node/.style={circle,fill=green,minimum size=0.75cm,inner sep=0pt}}
    \tikzset{two node/.style={circle,fill=magenta,minimum size=0.75cm,inner sep=0pt}}
    \tikzset{chiral node/.style={circle,fill=cyan,minimum size=0.75cm,inner sep=0pt}}
    \tikzset{constructive node/.style={circle,fill=yellow,minimum size=0.75cm,inner sep=0pt}}
    \tikzset{f node/.style={circle,fill=orange,minimum size=0.75cm,inner sep=0pt}}
   \tikzset{name node/.style={circle,fill=white,minimum size=0.75cm,inner sep=0pt}}
      \node[root node] (1) {$\ket{e_5}$};
      \node[two node] (2) [left = 0.5cm of 1]  {$\ket{e_4}$};
      \node[chiral node] (3) [left = 1.875cm of 1]  {$\ket{e_3}$};
      \node[constructive node] (4) [left = 3.25cm of 1]  {$\ket{e_2}$};
      \node[f node] (5) [left = 4.5cm of 1]  {$\ket{e_1}$};
      \path[draw=green,thick]
      (1) edge node {} (2);
      \path[draw=magenta,thick]
      (2) edge node {} (3);
      \path[draw=cyan,thick]
      (3) edge node {} (4);
      \path[draw=yellow,thick]
      (4) edge node {} (5);
    \end{tikzpicture}
    
    \caption{(Upper panel): Effective topology for the dynamics of an \textit{input} state. The \textit{input nodes} ($\ket{1}$ and $\ket{2}$) are green and magenta labeled, the \textit{chiral layer} ($\ket{k}$), composed of $d$ nodes, is blue labeled, the \textit{routing vertex} ($\ket{r}$) is yellow labeled and the output $\ket{f}$ is orange labeled. (Lower panel): Krylov representation of the effective topology. Following the same color palette the Krylov \textit{input vectors} $\ket{e_{5}}$ and $\ket{e_{4}}$ are green and magenta labeled, the Krylov \textit{chiral vector} $\ket{e_{3}}$ is blue labeled, the Krylov \textit{routing vector} $\ket{e_{2}}$ is yellow labeled and the Krylov \textit{output vector} $\ket{e_{1}}$ is orange labeled}
    \label{fig:Effective Lily}
\end{figure}

The reduced Hamiltonian in the Krylov basis is always five-dimensional, 
and  depends on the chiral layer only through the dimension $d$, see Fig.\ref{fig:Effective Lily}. Upon evaluating the elements
 $\bra{e_{j}}  \widetilde{H} \ket{e_{k}}$ we have the matrix representation 
 \begin{equation}
	\label{eq:reduced_Hamiltonian_beta}
	 \widetilde{H}  =
\begin{pmatrix}
 	0 & 1 & 0 & 0 & 0 \\
	1 & 0 & \sqrt{d} & 0 & 0 \\
	0 & \sqrt{d} & 0 & \sqrt{d} & 0 \\
	0 & 0 & \sqrt{d} & 0 & 1  \\
    0 & 0 & 0 & 1 & 0 & \\
\end{pmatrix}\,.
\end{equation}

This means that the time evolution does not depend on the overall 
dimensionality of the graph, and thus the routing procedure is 
associated to a universal time $t^{*}$ valid for every number 
of output spatial regions $n$ and then, for every $\ket{\psi_{f}}$.

\begin{figure}[!ht]
    \centering
    \includegraphics[width=\linewidth]{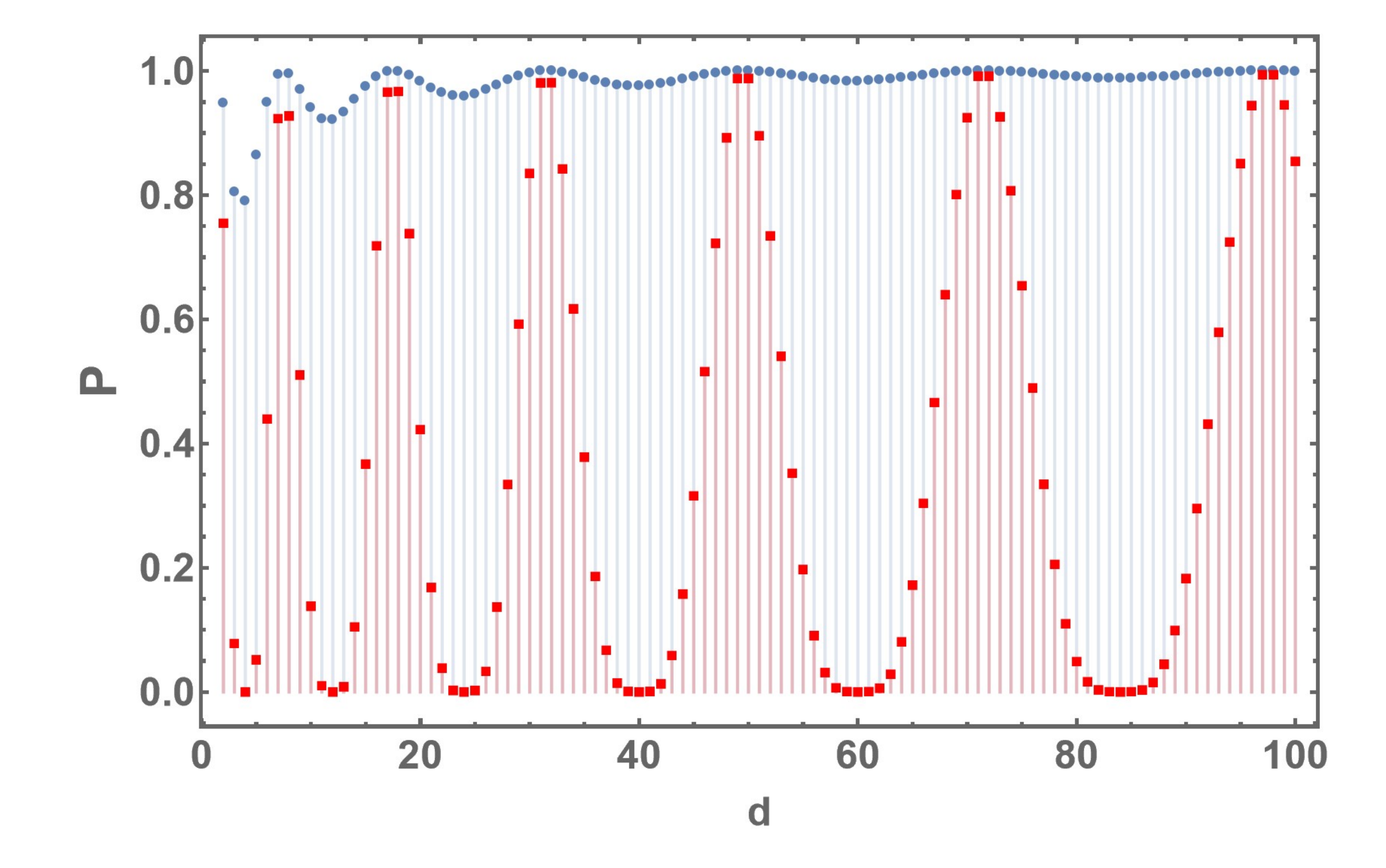}
    \caption{Routing probabilities $P_{1f}(\pi,d)$ (grey dots) and $P_{2r}(\pi,d)$ (red squares) as functions of the number of elements in the routing layer, $d$. $P_{1f}(\pi,d)$ slightly depends on $d$, being always really close to 1, while $P_{2r}(\pi,d)$ shows a sequence of maxima progressively closer to 1.}
    \label{fig:imp}
\end{figure}

The two probabilities $ P _{1f}(t)=\left|\langle e_1|e^{-i t \widetilde{H}}|e_5\rangle \right|^2$ and $ P _{2r}(t)=\left|\langle e_2|e^{-i t \widetilde{H}}|e_4\rangle \right|^2$ govern the routing performance for localized states. Using Eq. (\ref{eq:reduced_Hamiltonian_beta}), we have

\begin{align}
    \label{eq:probabilities}
         P _{1f}(t) &= \frac{
         \left[ 2d-(2d+1) \cos(t)+\cos\left(t \sqrt{1+2d} \right) \right]^2}
        {4(1+2d)^2}\notag
        \\ 
         P _{2r}(t) &= \frac14 \left[ \cos{(t)} - \cos{\left(t \sqrt{1+2d}\right)} \right]^{2}\, .
\end{align}
Upon setting $t^* \simeq \pi$, we have $ P _{1f}(t^{*}) \simeq 1$, with 
$$ P _{1f}(t^{*}) = 1 - \frac{1- \cos (\pi\sqrt{2d+1})}{2d} +O\left(\frac{1}{d^2}\right)\,,$$  which is approaching $1$ for increasing $d$, and $P_{2r}(t)$ with a sequence of maxima also progressively closer to 1, despite never being exactly 1 (see Fig.\ref{fig:imp}). Then, the chiral Lily graph provides nearly perfect routing of classical information (localized states), especially for large $d$, in a time independent of the number of outputs $n$. 

\begin{figure}[!ht]
\includegraphics[width=0.95\columnwidth]{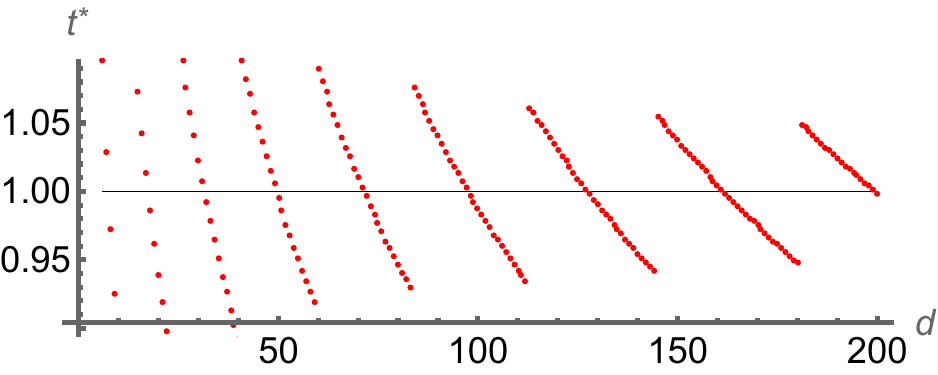}
\includegraphics[width=0.95\columnwidth]{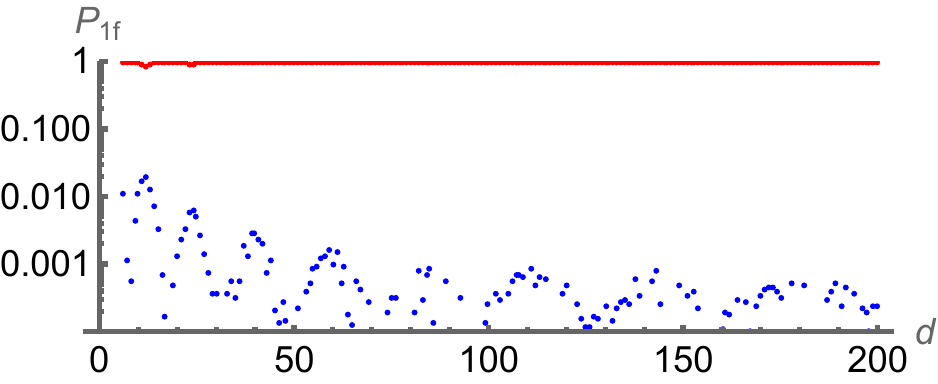}
\caption{(Upper panel): optimal time $t^*$ (in unit of $\pi$) maximizing the routing probability $P _{1f}(t)$ as a function of the chiral layer dimension $d$. (Lower panel): log-plot of the maximized routing probability $P _{1f}(t^*)$ as a function of the chiral layer dimension $d$. We also show the difference $|P _{1f}(t^*)-P_{2r}(t^*)|$ (the curve is below $10^{-2}$ as far as $d$ is larger than 15).
\label{npr}}
\end{figure}

With a finer analysis it is possible to define an optimal local routing time, around $\pi$, specific for every $d$ (see upper panel of Fig.\ref{npr}). Upon selecting this precise $t^{*}$ which maximizes $P _{1f}(t)$, it is possible to obtain an almost unitary fidelity for every dimensionality of the routing layer. Additionally, this value $t^{*}$, turns out to be a sound choice also for the maximization $P _{2r}(t)$, which assumes a value close to $P _{1f}(t)$ independently of $d$, and then almost unitary for every configuration of the Lily Graph (see lower panel of Fig.\ref{npr}).

$ P _{1f}(t^{*}) =  P _{2r}(t^{*}) = 1$ is also a necessary, but not sufficient, condition to obtain perfect routing also of any superposition of the states $\ket{e_5}$ and $\ket{e_4}$. The sufficient condition is

\begin{equation}
\label{eq:sufficient}
    \langle e_1|e^{-i t^* \widetilde{H}}|e_5\rangle=\langle e_1|e^{-i t^* \widetilde{H}}|e_5\rangle=e^{i\varphi}
\end{equation}
for any $\varphi \in [0,2\pi) $, and asks for the additional requirement that the phase acquired by $\ket{e_5}$ during the evolution must be the same as the phase acquired by $\ket{e_4}$. The quantity
\begin{equation}
    \Delta_{\varphi}(t,d)\equiv Arg[\langle e_1|e^{-i t \widetilde{H}}|e_5\rangle]-Arg[\langle e_2|e^{-i t \widetilde{H}}|e_4\rangle]
\end{equation} 
where $Arg[.]$ denotes the phase of a complex number and intrinsically defines the validity of the protocol with respect to the phase $\gamma$ of the initial and final routing state. From the matrix exponentiation of Eq.\eqref{eq:reduced_Hamiltonian_beta}, the transition amplitudes $\langle e_1|e^{-i t \widetilde{H}}|e_5\rangle$ and $\langle e_2|e^{-i t \widetilde{H}}|e_4\rangle$ turn out to be purely real, which means that $\Delta_{\varphi}(t,d)$ can be either $0$ or $\pi$. In particular, selecting $t=\pi$ we obtain $\Delta_{\varphi}(\pi,d)=0, \forall d$. This implies a direct correlation between high localized transition probabilities (like the ones observed in Fig.\ref{fig:imp}) and a high transition probability also for a superposition state like the one in Eq.\eqref{eq:input_state}. We are then able to find many configurations achieving nearly perfect routing of both classical (localized states) and quantum (superpositions) information, for specific $d$, in a time independent of the number of outputs, the form of the initial state and of the dimension of the routing layer.

\section{Perfect routing using chirality and edge weighting}
\label{sec:PRUCAEW}
Perfect routing may be obtained by slightly modifying  the Hamiltonian in Eq.\eqref{eq:Lily_Hamiltonian}, tuning the weights of the edges in the chiral and routing layers of the Lily graph as follows 
\begin{align}
    \label{eq:weighted_Lily_Hamiltonian}
     H  \left( n,d,\vec{\phi} \right) = A_{\mathcal{I}} + \beta\, C_{\mathcal{C}} \left( d,\vec{\phi} \right) + \beta\, C_{\mathcal{R}} \left( d,\vec{\phi},n \right) +A_{\mathcal{O}} ( n )\,,
\end{align}
The Krylov basis is the same of Eq.\eqref{eq:Krylov Basis}, 
and the reduced  Hamiltonian reads
\begin{equation}
	\label{eq:reduced_Hamiltonian}
	 \widetilde{H}(\beta) =
\begin{pmatrix}
 	0 & 1 & 0 & 0 & 0 \\
	1 & 0 & \beta \sqrt{d} & 0 & 0 \\
	0 & \beta \sqrt{d} & 0 & \beta \sqrt{d} & 0 \\
	0 & 0 & \beta \sqrt{d} & 0 & 1  \\
    0 & 0 & 0 & 1 & 0 & \\
\end{pmatrix},
\end{equation}
The routing probabilities become
\begin{align}
    \label{eq:probabilities_beta}
         P _{1f}(\beta,t) & = \frac{\left[ 2d\beta^{2}-(2d+\beta^{2})\cos{(t)+\cos{\left(t \sqrt{\beta^{2}+2d}\right)}} \right]^2}{4(\beta^{2}+2d)^2} \notag \\
         P _{2r}(\beta,t) & = \frac14 \left[ \cos{(t)} - \cos{\left(t \sqrt{\beta^{2}+2d} \right)} \right]^{2}\, .
\end{align}
By choosing $\beta \sqrt{d} = \sqrt{3/2}$ \footnote{Or any $\beta$ solution of the equation $\sqrt{2\beta^2 + 1} = 2 q,$ with $q \in \mathbb{N}$}, we achieve perfect routing for any classical state localized in the input region, for $t^{*}_{\beta}=\pi$. The condition of unit probability can be obtained also from the analysis of the energy level of the Krylov Hamiltonian. $\Tilde{\mathcal{H}}$ results to be the spin rotation matrix $S_{x}$ associated to a spin $s=2$. Its spectrum is composed of equally distant eigenvalues, which are proven to be a useful resource in quantum transport problems \cite{liu2023designing}. The simplest realization of the Lily graph is obtained by setting $d=2$, which requires only two phases $\phi=0,\pi$ and setting $\beta=\sqrt{3}/2$. However, the routing time is universal, i.e. independent on $n$ and $d$. Upon expanding 
the probabilities for times around $t^*$, we have

\begin{align}
P _{1f}(t) & = 1 - (t-t^*_\beta)^2 + O(t-t^*_\beta)^3 \notag \\ 
P _{2r}(t) & = 1 - \frac52 (t-t^*_\beta)^2 + O(t-t^*_\beta)^3 \,, \notag
\end{align}
$\forall n,d$, i.e., the routing is robust against fluctuations, and the robustness is universal too.

Associated to perfect classical routing, the system also shows a temporal periodicity of $2\pi$, assuring that for every time $t^{*}_{\beta}=(2q+1)\pi$ the projector condition 

\begin{align}
    \label{eq:Projector}
    e^{-i  \widetilde{H} (2q+1)\pi} = \sum_{s=0}^{4} \ket{e_{1+s}}\bra{e_{5-s}}, \;\; q \in \mathbb{N}\,,
\end{align}
is fulfilled, leading to a universal routing time for any input state (classical or quantum), independently of its form, as

\begin{align}
    \label{eq:routing}
    \ket{\psi_{f}} &= e^{-i  \widetilde{H} (2q+1)\pi} \left( \alpha\ket{e_{5}} + e^{i\gamma}\sqrt{1-\alpha^{2}}\ket{e_{4}} \right) = \nonumber \\
    &=\alpha\ket{e_{1}} + e^{i\gamma}\sqrt{1-\alpha^{2}}\ket{e_{2}},
\end{align}
fulfilling the condition of Eq.\eqref{eq:sufficient}. Notice that the coherence (in the site basis) of the evolving states 
$e^{-i t \widetilde{H}}|e_4\rangle$ and $e^{-i t \widetilde{H}}|e_5\rangle$ is instead characterized by a period of $\pi$. If $\beta \sqrt{d} \neq \sqrt{3/2}$ this periodicity is lost.

\section{Qudit Routing} 
\label{sec:QR}
Once the routing problem was addressed for a classical bit and a qubit, it can be extended to larger systems to define the routing of a general superposition of $\Tilde{D}$ reticular sites (i.e. a qudit) \footnote{This extension paves the way not only to the qudit routing, but also to the qubit routing over larger distances.}. 

\begin{figure}[h!] 
    \begin{tikzpicture}[thick,scale=0.7, every node/.style={scale=0.7}]
    \tikzset{root node/.style={circle,fill=green,minimum size=0.75cm,inner sep=0pt}}
    \tikzset{two node/.style={circle,fill=magenta,minimum size=0.75cm,inner sep=0pt}}
    \tikzset{chiral node/.style={circle,fill=cyan,minimum size=0.75cm,inner sep=0pt}}
    \tikzset{constructive node/.style={circle,fill=yellow,minimum size=0.75cm,inner sep=0pt}}
    \tikzset{f node/.style={circle,fill=orange,minimum size=0.75cm,inner sep=0pt}}
    
    \tikzset{dots node/.style={circle,fill=white!40,minimum size=0.75cm,inner sep=0pt}}
    
      \node[f node] (5) {$f_{\Tilde{D}+1}$};

      \node[f node] (5bis) [above=0.75cm of 5] {$f_{2\Tilde{D}+1}$};
      \node[f node] (6bis) [right=3.75cm of 5bis] {$o_{2\Tilde{2}+1}$};
      \node[f node] (7bis) [left=3.75cm of 5bis] {$o_{2\Tilde{2}+1}$};
      
      \node[constructive node] (4) [below = 0.75cm of 5]  {$r$};
      
      \node[chiral node] (3) [below = 2.125cm of 5]  {$k$};
      \node[two node] (2) [below = 3.25cm of 5]  {$\Tilde{D}$};
      \node[two node] (1) [below = 4.5cm of 5]  {$2$};
      \node[root node] (1bis) [below = 5.625cm of 5]  {$1$};
      \node[chiral node] (6) [left = 0.75cm of 3]  {$k$};
      \node[chiral node] (8) [right = 0.75cm of 3]  {$k$};
      \node[constructive node] (10) [left = 2.0cm of 4]  {$l$};     
      \node[constructive node] (11) [right = 2.0cm of 4]  {$l$}; 
      \node[f node] (12) [left = 2.9cm of 5]  {$o_{\Tilde{D}+3}$};     
      \node[f node] (13) [right = 2.9cm of 5]  {$o_{\Tilde{D}+3}$}; 
      \path[draw=black,dotted]
      (1) edge node {} (2);
      \path[draw=green,thick]
      (1bis) edge node {} (1);
      \path[draw=magenta,thick]
      (2) edge node {} (3)
      (2) edge node {} (6)
      (2) edge node {} (8);
      \path[draw=cyan,thick]
      (3) edge node {} (4)
      (4) edge node {} (6)
      (4) edge node {} (8);
      \path[draw=yellow,thick]
      (4) edge node {} (5); 
      \path[draw,dotted]
      (10) edge node {} (6)
      (10) edge node {} (3)
      (10) edge node {} (8);
      \path[draw,dotted]
      (11) edge node {} (6)
      (11) edge node {} (3)
      (11) edge node {} (8);
      \path[draw,dotted]
      (10) edge node {} (12)
      (11) edge node {} (13)
      (5) edge node {} (5bis)
      (13) edge node {} (6bis)
      (12) edge node {} (7bis);
    \end{tikzpicture}
    \caption{\textit{Generalized Lily Graph} topology -- The \textit{input nodes} ($\ket{1}$, $\ket{2}$ up to $\Tilde{\ket{D}}$) are green and magenta labeled, the \textit{chiral layer} ($\ket{k}$), composed of $d$ nodes, is blue labeled, the \textit{routing layer} ($\ket{l}$ and $\ket{r}$), composed of $n$ vertices is yellow labeled and the $n$ outputs, $\ket{f_{l}}$ and $\ket{o_{l}}$, are orange labeled.}
    \label{fig:Generalized Lily Graph}
\end{figure}

Starting from the basic Hamiltonian for qudit routing, i.e. Eq.\eqref{eq:weighted_Lily_Hamiltonian}, we can generalize the routing process for a $\Tilde{D}$-dimensional input region, defined by the $A^{\Tilde{D}} \left( \Tilde{D} \right)$ adjacency, and a $\Tilde{D}$-dimensional output region, defined by the $A^{\Tilde{D}}_\mathcal{O} ( \Tilde{D},n )$, as 

\begin{align}
    \label{eq:General_weighted_Lily_Hamiltonian}
     H  \left( n,d,\vec{\phi},\Tilde{D}  \right) & =  A^{\Tilde{D}}_{\mathcal{I}} \left( \Tilde{D} \right) + C_{\mathcal{C}} \left( d,\vec{\phi} \right) + \nonumber \\
     & C_{\mathcal{R}} \left( d,\vec{\phi},n \right) +A^{\Tilde{D}}_\mathcal{O} ( \Tilde{D},n ).
\end{align}
Each element of the Hamiltonian must be defined as follow,

\begin{align}
    \label{eq:General_weighted_input_adjacency}
     A^{\Tilde{D}}_{\mathcal{I}} \left( \Tilde{D} \right) = \frac{1}{2} \sum_{l=1}^{\Tilde{D}-1} &\sqrt{l(2\Tilde{D}+1-l)} \ket{l} \bra{l+1} \nonumber \\
     &+ h.c.,
\end{align}

\begin{align}
    \label{eq:General_weighted_chiral_adjacency}
     C_{\mathcal{C}} \left( d,\vec{\phi} \right) = \frac{1}{2}\sqrt{\frac{\Tilde{D}(\Tilde{D}+1)}{d}} \bigg(\sum_{k \in \mathcal{C}} & e^{-i\phi_{k}} \Tilde{\ket{D}} \bra{k} + \nonumber \\
     & e^{i\phi_{k}} \ket{k}\Tilde{\bra{D}} \bigg),
\end{align}

\begin{align}
    \label{eq:General_weighted_routing_adjacency}
     C_{\mathcal{R}} \left( d,\vec{\phi},n \right) = &  \frac{1}{2}\sqrt{\frac{\Tilde{D}(\Tilde{D}+1)}{d}} \bigg( \sum_{\substack{l \in \mathcal{R} \\ l \neq r}} \, \sum_{k \in \mathcal{C}}  \ket{k} \bra{l} + \ket{l}\bra{k} \nonumber \\
    +& \sum_{k \in \mathcal{C}} e^{-i\phi_{k}} \ket{r}\bra{k} + e^{i\phi_{k}} \ket{k}\bra{r} \bigg),
\end{align}

\begin{align}
    \label{eq:General_weighted_output_adjacency}
     A^{\Tilde{D}}_{\mathcal{O}} \left( \Tilde{D} \right) = &\frac{1}{2} \sqrt{(\Tilde{D}+2)(\Tilde{D}-1)} \sum_{l \in \mathcal{R}} \sum_{a \in \mathcal{O}} \ket{a} \bra{l} + \nonumber \\ & \frac{1}{2} \sum_{\substack{b \in \mathcal{O} \\ l=\Tilde{D}+3}}^{2\Tilde{D}+1} \sqrt{l(2\Tilde{D}+1-l)} \ket{b_l} \bra{b_{l+1}} \nonumber \\
     &+ h.c.,,
\end{align}

The phase distributions in $ C_{\mathcal{C}}$ and $ C_{\mathcal{R}}$ are the same defined in Eqs. \eqref{eq:deph_adjacency} and \eqref{eq:int_adjacency}. 
Following the same Krylov dimensionality reduction method applied for the qubit routing (Eqs.\eqref{eq:dim_red}-\eqref{eq:Gram_Schmidt}) the reduced basis turns out to be

\begin{align}
    \label{eq:Generalized Krylov Basis}
    &\ket{e_{1}}= \ket{f_{2\Tilde{D}+1}}\,, \quad \ket{e_{2}}=\ket{f_{2\Tilde{D}}}\nonumber \\
    & \quad \quad \quad \quad \quad \quad \quad \vdots \nonumber \\
    &\ket{e_{\Tilde{D}-1}}= \ket{f_{1}}\,, \quad
    \ket{e_{\Tilde{D}}}= \ket{r} \nonumber \\
    &\ket{e_{\Tilde{D}+1}}= \frac{1}{\sqrt{d}}\sum_{k \in \mathcal{C}} e^{i\phi_{k}} \ket{k} \nonumber \\
    &\ket{e_{\Tilde{D}}+2}= \ket{\Tilde{D}}\,, \quad
    \ket{e_{\Tilde{D}+3}}= \ket{\Tilde{D}-1} \nonumber \\
    & \quad \quad \quad \quad \quad \quad \quad \vdots \nonumber \\
    &\ket{e_{2\Tilde{D}}}= \ket{2} \,, \quad
    \ket{e_{2\Tilde{D}+1}}= \ket{1}
\end{align}

The weight distribution over the edges of the Hamiltonian described in Eqs. \eqref{eq:General_weighted_input_adjacency}-\eqref{eq:General_weighted_output_adjacency} ensures that in the Krylov reduced basis the Hamiltonian reads as

\begin{equation}
	\label{eq:General_reduced_Hamiltonian}
	 \widetilde{H}(\Tilde{D}) = S^{2\Tilde{D}+1}_{x},
\end{equation}
where the $S^{2\Tilde{D}+1}_{x}$ is the spin rotation matrix \cite{christandl2004perfect,christandl2005perfect}, and the apex $2\Tilde{D}+1$ refer to its dimension \footnote{The dimension of the Krylov reduced Hamiltonian for a Qudit has to be $ 2\Tilde{D}+1 $, because of the $\Tilde{D}$ inputs, the $\Tilde{D}$ outputs and the central layer.}. Such structure of the Hamiltonian ensure that at a time $t^{*}=(2q+1)\pi$ the time evolution operator results

\begin{align}
    \label{eq:General_Projector}
    e^{-i  \widetilde{H}(\Tilde{D}) (2q+1)\pi} = \sum_{s=0}^{2\Tilde{D}+1} \ket{e_{1+s}}\bra{e_{2\Tilde{D}+1-s}}, \;\; q \in \mathbb{N}\,,
\end{align}
allowing perfect periodical chiral quantum routing of any superposition of input state to the selected output region. 

\begin{figure}[!ht] 
    \begin{tikzpicture}[thick,scale=0.9, every node/.style={scale=0.9}]
    \tikzset{root node/.style={circle,fill=green,minimum size=0.75cm,inner sep=0pt}}
    \tikzset{two node/.style={circle,fill=magenta,minimum size=0.75cm,inner sep=0pt}}
    \tikzset{chiral node/.style={circle,fill=cyan,minimum size=0.75cm,inner sep=0pt}}
    \tikzset{constructive node/.style={circle,fill=yellow,minimum size=0.75cm,inner sep=0pt}}
    \tikzset{f node/.style={circle,fill=orange,minimum size=0.75cm,inner sep=0pt}}
    
    \tikzset{dots node/.style={circle,fill=white!40,minimum size=0.75cm,inner sep=0pt}}

    \tikzset{name node/.style={circle,fill=white,minimum size=0.75cm,inner sep=0pt}}
    
      \node[f node] (5) {$\scriptstyle f_{2\Tilde{D}+1}$};
      
      \node[f node] (f-1) [below = 0.50cm of 5] {$f_{1}$};
      \node[constructive node] (4) [below = 2cm of 5]  {$r$};
      \node[chiral node] (3) [below left = 3.35cm and 0.5cm of 5]  {$k$};
      \node[two node] (d) [below = 4.25cm of 5]  {$\Tilde{D}$};
      \node[two node] (2) [below = 5.625cm of 5]  {$2$};
      \node[root node] (1) [below = 6.625cm of 5]  {$1$};

      \node[chiral node] (6) [left = 0.75cm of 3]  {$k$};
      \node[chiral node] (7) [right = 1.25cm of 3]  {$k$};
      \node[chiral node] (8) [right = 0.75cm of 7]  {$k$};

      \node[dots node] (9) [right = 0.25cm of 3] {$...$};


      \path[draw=green,thick]
      (1) edge node {} (2);
      \path[draw=magenta,thick]
      (d) edge node {} (3)
      (d) edge node {} (6)
      (d) edge node {} (7)
      (d) edge node {} (8);
      \path[draw=cyan,thick]
      (3) edge node {} (4)
      (4) edge node {} (6)
      (4) edge node {} (7)
      (4) edge node {} (8);
      \path[draw=yellow,thick]
      (4) edge node {} (f-1);
      \path[draw=black,dotted]
      (2) edge node {} (d);
      \path[draw=black,dotted]
      (5) edge node {} (f-1);
    \end{tikzpicture}
    \bigskip

    \begin{tikzpicture}[thick,scale=0.9, every node/.style={scale=0.8}]
    \tikzset{root node/.style={circle,fill=green,minimum size=0.9cm,inner sep=0pt}}
    \tikzset{two node/.style={circle,fill=magenta,minimum size=0.9cm,inner sep=0pt}}
    \tikzset{chiral node/.style={circle,fill=cyan,minimum size=0.9cm,inner sep=0pt}}
    \tikzset{constructive node/.style={circle,fill=yellow,minimum size=0.9cm,inner sep=0pt}}
    \tikzset{f node/.style={circle,fill=orange,minimum size=0.9cm,inner sep=0pt}}
   \tikzset{name node/.style={circle,fill=white,minimum size=0.9cm,inner sep=0pt}}
      \node[root node] (1) {$\ket{e_{ {\scalebox{0.5}{$2\Tilde{D} + 1$}}}}$};
      \node[two node] (2) [left = 0.5cm of 1]  {$\ket{e_{ 2\Tilde{D}}}$};
      \node[chiral node] (3) [left = 1.875cm of 1]  {$\ket{e_{\Tilde{D}}}$};
      \node[constructive node] (4) [left = 3.25cm of 1]  {$\ket{e_{ {\scalebox{0.5}{$\Tilde{D} + 1$}}}}$};
      \node[f node] (5) [left = 4.6cm of 1]  {$\ket{e_1}$};
      \path[draw=green,thick]
      (1) edge node {} (2);
      \path[draw=magenta,dotted]
      (2) edge node {} (3);
      \path[draw=cyan,thick]
      (3) edge node {} (4);
      \path[draw=black,dotted]
      (4) edge node {} (5);
    \end{tikzpicture}
    
    \caption{(Upper panel): Effective topology for the dynamics of an \textit{input} state for the \textit{Generalized Lily Topology}. The \textit{input nodes} ($\ket{1}$, $\ket{2}$ up to $\Tilde{D}$) are green and magenta labeled, the \textit{chiral layer} ($\ket{k}$), composed of $d$ nodes, is blue labeled, the \textit{routing vertex} ($\ket{r}$) is yellow labeled and the output $\ket{f}$ is orange labeled. (Lower panel): Krylov representation of the effective topology. Following the same color palette the Krylov \textit{input vectors} $\ket{e_{5}}$ and $\ket{e_{4}}$ are green and magenta labeled, the Krylov \textit{chiral vector} $\ket{e_{3}}$ is blue labeled, the Krylov \textit{routing vector} $\ket{e_{2}}$ is yellow labeled and the Krylov \textit{output vector} $\ket{e_{1}}$ is orange labeled}
    \label{fig:Generalized Effective Lily}
\end{figure}

Any normalized quantum state which correspond to a superposition of input nodes (nodes from $\ket{1}$ to $\Tilde{\ket{D}}$) in the form

\begin{equation}
    \label{eq:input_state_qudit}
    \ket{\psi^{\Tilde{D}}_{0}} = \sum_{l=1}^{\Tilde{D}}\alpha_{l} \ket{e_{2\Tilde{D}+2-l}} 
\end{equation}
will be routed to the desired output (nodes $\ket{r}$ and $\ket{f_{\Tilde{D}}}$, ..., $\ket{f_{2\Tilde{D}+1}}$ ) as

\begin{align}
    \label{eq:output_state_qudit}
    \ket{\psi^{\Tilde{D}}_{f}} &= e^{-i  \widetilde{H}(\Tilde{D}) (2q+1)\pi} \ket{\psi^{\Tilde{D}}_{0}} = \nonumber \\
    &= \left( \sum_{l=1}^{\Tilde{D}-1}\alpha_{l} \ket{e_l} \right) + \alpha_{\Tilde{D}}\ket{e_{\Tilde{D}}}.
\end{align}
The time required to perform such procedure results again independent of the form of the input state, conferring a universal validity to the scheme.

\section{Physical Implementations} 
\label{sec:PI}
After the analytical description of the Hamiltonian and its properties, in this section we address the possible physical implementations of such protocol. Consider a time-independent Hamiltonian of a Heisenberg spin structure (Fig.\ref{fig:Generalized Lily Graph}) in the form

\begin{align}
    \label{eq:spin_chain}
    H \propto \frac{1}{2} \sum_{j \in \textit{Lily}} J_{j} \vec{\sigma}_{j} \cdot \vec{\sigma}_{j+1} + \sum_{j \in \textit{Lily}} B_{j} \sigma_{j}^{z}
\end{align}
which in the reduced Krylov space (Fig.\ref{fig:Generalized Effective Lily}) leads to

\begin{align}
    \label{eq:spin_chain_reduced}
    \widetilde{H} \propto \frac{1}{2} \sum_{e_j \in \textit{Krylov}} \widetilde{J}_{e_j} \vec{\sigma}_{e_j} \cdot \vec{\sigma}_{e_{j+1}} + \sum_{e_j \in \textit{Krylov}} \widetilde{B}_{e_{j}} \sigma_{e_j}^{z}.
\end{align}
If the $\widetilde{J}_{e_j}$ and $\widetilde{B}_{e_j}$ satisfy the relations

\begin{equation}
\label{eq:conditions_implementations}
\begin{cases}
\widetilde{J}_{e_j} = \sqrt{j(2\Tilde{D}+1-j)} \, , \\
\widetilde{B}_{e_j} = \frac{\left( \widetilde{J}_{e_{j-1}}+\widetilde{J}_{e_j} \right)}{2} - \frac{1}{2\left( 2\Tilde{D} -1\right)} \sum_{l=1}^{2\Tilde{D}}\widetilde{J}_{e_l} \, ,
\end{cases}
\end{equation}
which corresponds to Eq.\eqref{eq:General_weighted_Lily_Hamiltonian} in single particle regime, the system support perfect periodic chiral quantum routing. Such systems are valuable candidate for the implementation of routing devices. This means that in principle any spin structure with an appropriate application of position dependent magnetic field \cite{paganelli2013routing}, or any physical system with the same Hamiltonian \cite{perez2013perfect,chen2020efficient}, can be a suitable realization of a physical structure for quantum routing. 

\section{Conclusions} 
\label{sec:C}
In this work, we have focused on designing a device capable of routing quantum information, initially encoded on two input nodes of a network, to $n$ possible pairs of output nodes, which are orthogonal and mutually exclusive. Upon exploiting the formalism of continuous-time quantum walks, we have designed an optimal five-layer structure, named the Lily graph, comprising two input nodes, a chiral layer, a routing layer, and an output layer. The system's Hamiltonian has been engineered by leveraging both the chiral properties available for a quantum walker and the modulation of the graph's edge weights. The time evolution ensures a perfect routing protocol, achieving unit fidelity, for both classical information (initially localized at a graph site) and quantum information (initially encoded in a superposition of sites). The selection of output nodes is achieved by modifying the phases of certain links in the routing layer. The time required for information transfer is universal, i.e., it does not depend on the initial state, on the number of outputs or the size of the chiral layer, and the overall protocol is robust, with the output fidelity being relatively unaffected by fluctuations in the interaction time. The process can be extended to qudit routing (i.e. routing of any arbitrary quantum state, even over large distances), and can be implemented in spin structures with the application of a suitable magnetic field. Our results demonstrate that it is possible to achieve perfect routing of  quantum information on a network characterized by a relative simplicity,  and pave the way for future developments, including the routing of entanglement. The Lily graph structure provides a robust and scalable  method for quantum information routing, possibly enhancing the prospects for integrating quantum technologies into existing communication and computation infrastructures.

\begin{acknowledgments}
This work has been done under the auspices of GNFM-INdAM and has been partially supported by MUR through the project PRIN22-2022T25TR3-RISQUE and by MUR and EU through the project PRIN22-PNRR-P202222WBL-QWEST.
\end{acknowledgments}

\appendix

\bibliography{pcqrbib.bib}

\end{document}